\documentclass[twoside]{article}
\usepackage{fleqn,espcrc2}

\newcommand{\AmS}{{\protect\the\textfont2
  A\kern-.1667em\lower.5ex\hbox{M}\kern-.125emS}}
\newcommand{\be}{\begin{eqnarray}}
\newcommand{\ee}{\end{eqnarray}}
\newcommand{\bc}{\begin{center}}
\newcommand{\ec}{\end{center}}
\newcommand{\barl}{\begin{array}{rl}}
\newcommand{\barr}{\begin{array}{rr}}
\newcommand{\ball}{\begin{array}{llllll}}
\newcommand{\ea}{\end{array}}
\newcommand{\nnb}{\nonumber}
\newcommand{\bea}{\begin{eqnarray}}
\newcommand{\eea}{\end{eqnarray}}
\def\ml{{\hat{m_{\ell}}}}
\def\ddp{{D^\prime}}
\def\mc{{\hat{m_c}}}
\def\c{C}
\def\cs{{\c_7}}
\def\cn{{\c_9}}
\def\ct{{\c_{10}}}
\def\cne{\cn^{\rm eff}}
\def\cse{\cs^{\rm eff}}
\def\m{{\cal M}}

\def\g{\gamma}
\def\l{\ell}
\def\ks{{K^\ast}}
\def\lb{\bar{\l}}

\def\bxsll{$B \rightarrow X_s \ell^+ \ell^- $}
\def\bxqll{$B \rightarrow X_q \ell^+ \ell^- $}

\def\bkll{B \rightarrow K \, \l^+ \, \l^-}
\def\bksll{B \rightarrow \ks \, \l^+ \, \l^-}

\def\he{{\cal H}_{\rm eff}}

\def\d{{\rm d}}

\def\mh{\hat{m}}
\def\mbh{\mh_b}
\def\msh{\mh_s}
\def\mph{\mh_K}
\def\mvh{\mh_{K^*}}

\def\mlh{\mh_\l}
\def\qh{\hat{q}}

\def\sh{\hat{s}}

\def\a{{\cal A}}
\def\s{{\cal S}}
\def\cqb{C_{Q_1}}

\def\ap{{A^\prime}}

\def\cp{{C^\prime}}

\def\uh{{\hat{u}}}
\def\lam{{\lambda}}

\def\be{\begin{eqnarray}}
\def\ee{\end{eqnarray}}
\def\ba{\begin{eqnarray}}
\def\ea{\end{eqnarray}}
\def\nnb{\nonumber}

\def\nnb{\nonumber}

\def\journal#1#2#3#4{{\it #1} {\bf #2} (#3) #4}

\def\pl{Phys. Lett.}
\def\np{Nucl. Phys.}

\def\pr{Phys. Rev.}

\def\prdn#1#2#3{{Phys.~Rev. D}~{\bf #1}, #3 (20#2)}

\def \lop{P_{\mathrm L}}
\def \trp{P_{\mathrm T}}
\def \nop{P_{\mathrm N}}

\title{Rare decays $B\rightarrow X_s l^+l^-$ and $B\rightarrow K^{(*)}l^+l^-$
       in SM and beyond}

\author{Chao-Shang Huang\address{Institute of Theoretical Physics, Academia 
        Sinica, 100080 Beijing, China}}

\begin{document}

\begin{abstract}
After a brief review of the inclusive and exclusive rare semileptonic decays of B mesons
in the standard model (SM), an overview of recent theoretical developments in
this field is given.
New physics effects on various observables such as branching ratio, backward-forward 
asymmetry, polarization of lepton and CP violation in the decays are analyzed in models 
beyond SM (supersymmetric models and two Higgs doublet models). 
\vspace{1pc}
\end{abstract}

\maketitle

\section{INTRODUCTION}

There are no flavor changing neutral currents (FCNC) at
tree-level in the standard model (SM). FCNC appear at loop-levels and
consequently offer a good place to test quantum effects of the
fundamental quantum field theory on which SM based. Furthermore, they are
very small at one loop-level due to the unitarity of Cabbibo-Kobayashi-Maskawa
 (CKM) matrix. In models beyond SM new particles 
may appear in the loop and have significant contributions to flavor
changing transitions. Therefore, FCNC interactions give an ideal place to
search for new physics. Any positive observation of FCNC couplings
deviated from that in SM would unambiguously signal the presence of new
physics. Searching for FCNC is clearly one of important goals of the next
generation of high energy colliders~\cite{pro}.

In this talk I shall review the recent developments of rare decays $B\rightarrow X_s l^+l^-$ and 
$B\rightarrow K^{(*)}l^+l^-$, one kind of FCNC processes, in SM and beyond (SUSY and 2HDM)
 with emphasis on the latter. In 2HDM or SUSY the couplings of neutral Higgs bosons (NHBs)
to down-type quarks or leptons are proportional to $\frac{m_f}{m_w}$ tan$\beta$ which leads
to significant effects on observables if tan$\beta$ is large. So we shall pay particular 
attention to the large tan$\beta$ case in the talk. I am sorry to say that some interesting 
works are not addressed due to the limited length of the paper.

\section{INCLUSIVE DECAYS $B\rightarrow X_s l^+l^-$}

\subsection{In SM}

The effective Hamiltonian relevant to $b\rightarrow s$ transition in SM is

\begin{equation}
        \he = -4 \frac{G_F} {\sqrt{2}}  V_{t s}^\ast  V_{tb}
              \sum_{i=1}^{10} \c_i(\mu)  O_i(\mu),  
 \label{eq:eh}
\end{equation}
Where operators $O_i$ and Wilson coefficients $C_i$ in leading logarithmic approximation (LLA) 
can be found in ref.\cite{bm}. 
The above Hamiltonian leads to the following free quark decay amplitude
\begin{eqnarray}
 &&       \m(b\to s\ell^+\ell^-) \nonumber \\
&& =\frac{G_F \alpha}{\sqrt{2}  \pi} \,
                V_{t s}^\ast V_{tb} \, \left\{
                  \cne  \left[ \bar{s}  \g_\mu  L  b \right] \,
                   \left[ \lb  \g^\mu  \l \right]  \right. \nonumber \\
  &&     \left.      + \ct  \left[ \bar{s}  \g_\mu  L  b \right] \,
                         \left[ \lb  \g^\mu  \g_5  \l \right]
                \right. \nonumber \\
        & & \left.
                - 2 \mbh  \cse  \left[ \bar{s}  i  \sigma_{\mu \nu}
                        \frac{\qh^{\nu}}{\sh}  R  b \right]
                        \left[ \lb  \g^\mu  \l \right]
                \right\} \; ,
        \label{eq:m}
\end{eqnarray}
where $L/R \equiv {(1 \mp \g_5)}/2$, $s = q^2$, $q=p_{+} +p_{-}$,
$\sh=s/m_B^2$, $\mbh=m_b/m_B$. We put $m_s/m_b = 0$, but keep
the leptons massive. In Eq. (\ref{eq:m}) $\cne$ can be found in, e.g., ref.\cite{ali1}.

The dilepton invariant mass spectrum (IMS) including power corrections
in the HQET approach in \bxqll decays can be written as:
\begin{eqnarray}
&& \frac{d {\cal{B}}}{d \s} = \frac{d {\cal{B}}^0}{d \s}
+\frac{d {\cal{B}}^{1/m_b^2}}{d \s}+
\frac{d {\cal{B}}^{1/q^2}}{d \s} \; ,
\end{eqnarray}
where the first term corresponds to the free quark decay $b\rightarrow s l^+l^-$,
 the second term accounts for the ${\cal{O}}(1/m_b^2)$ power corrections\cite{ali1},
and the last term accounts for the non-perturbative interaction of a virtual
$u\bar{u}$- and $c\bar{c}$-quark loop with soft gluons\cite{buc}.
 From Eq. (\ref{eq:m}), we obtain IMS for $b\rightarrow s l^+l^-$\cite{gri}
\begin{eqnarray}
&& \frac{{\rm d}\Gamma(B\rightarrow X_s l^{+}l^{-})}{{\rm d}\hat{s}}
 = {\cal B}_{sl}  \frac{{\alpha}^2}
 {4 \pi^2 f(\mc)\kappa(\mc) } \nonumber \\
&&  \times  (1-\hat{s})^2 (1-\frac{4t^2}{\hat{s}})^{1/2}
 \frac{|V_{tb}V_{ts}^{*}|^2}{|V_{cb}|^2} D(\hat{s}) ~,\nonumber \vspace{5mm}\\
&&D(\hat{s}) = |C_9^{eff}|^2(1+\frac{2t^2}{\hat{s}})(1+2\hat{s}) \nonumber \\
&&      + 4|C_7^{eff}|^2(1+ \frac{2t^2}{\hat{s}})(1+\frac{2}{\hat{s}}) \nonumber \\
&&       + |C_{10}|^2 [ ( 1 + 2\hat{s}) + \frac{2t^2}{\hat{s}}(1-4\hat{s})] \nonumber \\
&&      +12 {\rm Re}(C_7^{eff} C_{9}^{eff*})(1+\frac{2t^2}{\hat{s}}) 
\label{eq22}
\end{eqnarray}
where $t=m_l/m_b$.\\ 
The forward-backward asymmetry (FBA) for $b\rightarrow s l^+l^-$ is given by
\begin{eqnarray}
&& A(\hat{s})=\frac{\int^{1}_{0}dz \frac{d^2\Gamma}{d\hat{s} dz} -
\int^{0}_{-1}dz \frac{d^2\Gamma}{d\hat{s} dz}}{
\int^{1}_{0}dz \frac{d^2\Gamma}{d\hat{s} dz} +
\int^{0}_{-1}dz \frac{d^2\Gamma}{d\hat{s} dz}}
=\frac{E(\hat{s})}{D(\hat{s})},
\end{eqnarray}
\begin{eqnarray}
&& E(\hat{s})={\rm Re} (C_9^{eff} C_{10}^* \hat{s}+2 C_7^{eff} C_{10}^*),
\label{eq26}
\end{eqnarray}
where $z=\cos\theta$ and $\theta$ is the angle between the momentum
of B  and that of $l^+$ in the center of mass frame of the
dileptons $l^+l^-$. 
The longitudinal, transverse and normal polarizations of lepton (LP) are given by
\bea
&& P_L =  (1-\frac{4 t^2}{\hat{s}})^{1/2} \frac{D_L(\hat{s})}{D(\hat{s})},
\nonumber \\
&& P_N = \frac{3 \pi}{4 \hat{s}^{1/2}} (1-\frac{4 t^2}{\hat{s}})^{1/2}
\frac{D_N(\hat{s})}{D(\hat{s})},
\nonumber \\
&& P_T = -\frac{3 \pi t}{2 \hat{s}^{1/2}}
\frac{D_T(\hat{s})}{D(\hat{s})},
\eea
where
\bea
D_L(\hat{s}) &=& {\rm Re}\left(
 2 (1+2 \hat{s}) C_9^{eff} C_{10}^*+12 C_7^{eff} C_{10}^*  \right), \nonumber \\
D_N(\hat{s}) &=&  {\rm Im} \left(
4 t C_{10} C_7^{eff*}+ 2 t \hat{s} C_9^{eff\ *} C_{10} \right),
 \nonumber \\
D_T(\hat{s}) &=&   {\rm Re}\left(
-2 C_7^{eff} C_{10}^*+ 4 C_9^{eff} C_7^{eff*} +\frac{4}{\hat{s}} |C_7^{eff}|^2 \right. \nnb \\
&&\left. - C_9^{eff} C_{10}^* \right. 
\left. +\hat{s} |C_9^{eff}|^2 \right).
\label{eq30}
\eea

In LLA
one obtains the following numerical results \cite{ks,ali2}
\bea
&& Br(B\to X_s \, \mu^+ \mu^-) = 6.7\times 10^{-6}~, \nnb \\
&& Br(B\to X_s \, \tau^+ \tau^-) = 2.5\times 10^{-7}~.\label{eq:bsl} 
\eea
Apart from the CKM-parametric dependence (estimate $\pm 13\%$), the theoretical
uncertainty on the branching ratio comes mainly from the scale dependence. \\
The present experimental bound is\cite{cleo}
\bea
&& {\cal B}(B \to X_s \ell^+ \ell^-) < 4.2 \times 10^{-5} ~(at ~90\% ~C.L.) \nnb
\eea
Comparing the bound with Eq. (\ref{eq:bsl}), it follows that there is a room for new physics. 

For the lepton polarization, one has\cite{ks}
$\left\langle \lop \right\rangle_{\tau}
= -0.37$, $\left\langle \trp \right\rangle_{\tau} = -0.63$, $\left\langle \nop \right\rangle_{\tau} = 0.03\,
(0.02)$ for $\kappa_V= 2.35\,(1)$. 
Assuming a total of $5\times 10^8$ $B\bar{B}$ decays, one can expect to observe
$\sim 100$ identified $B\to X_s \, \tau^+ \tau^-$ events, permitting a test of the predicted polarization
$\left\langle \lop \right\rangle = -0.37$, and $\left\langle \trp \right\rangle = -0.63$ with good accuracy.
\subsection{Beyond SM}

We limit to discuss the model II 2HDM and SUSY (MSSM, mSUGRA, and string theory and M-theory inspired models) in the talk.
In these models one should include the contributions from exchanging NHBs in the large tan$\beta$ case. Instead of Eq. (\ref{eq:eh}),
the effective Hamiltonian describing $B\rightarrow X_s l^+l^-$ now becomes\cite{dhh}
\begin{eqnarray}\label{ham}
 H_{eff}&=&
\frac{4G_F}{\sqrt{2}}V_{tb}V^*_{ts}(\sum_{i=1}^{10}C_i(\mu)O_i(\mu)   \nnb
\\ && +\sum_{i=1}^{10}C_{Q_i}(\mu)Q_i(\mu)),
\end{eqnarray}
where $O_i$ is the same as that in Eq. (\ref{eq:eh}) and
$Q_i$'s come from exchanging neutral Higgs bosons and are defined by\cite{dhh}
\begin{eqnarray}\nonumber
Q_1&=&\frac{e^2}{16\pi^2}(\bar{s}^{\alpha}_Lb^{\alpha}_R)(\bar{\tau}
\tau)\\\nonumber
Q_2&=&\frac{e^2}{16\pi^2}(\bar{s}^{\alpha}_Lb^{\alpha}_R)(\bar{\tau}\gamma_5
\tau)\\\nonumber
Q_3&=&\frac{g^2}{16\pi^2}(\bar{s}^{\alpha}_Lb^{\alpha}_R)(\sum_q\bar{q}^{\beta}
_Lq^{\beta}_R)\\\nonumber
Q_4&=&\frac{g^2}{16\pi^2}(\bar{s}^{\alpha}_Lb^{\alpha}_R)(\sum_q\bar{q}^{\beta}
_Rq^{\beta}_L)\\\nonumber
Q_5&=&\frac{g^2}{16\pi^2}(\bar{s}^{\alpha}_Lb^{\beta}_R)(\sum_q\bar{q}^{\beta}
_Lq^{\alpha}_R)\\\nonumber
Q_6&=&\frac{g^2}{16\pi^2}(\bar{s}^{\alpha}_Lb^{\beta}_R)(\sum_q\bar{q}^{\beta}
_Rq^{\alpha}_L)\\\nonumber
Q_7&=&\frac{g^2}{16\pi^2}(\bar{s}^{\alpha}_L\sigma^{\mu\nu}b^{\alpha}_R)
(\sum_q\bar{q}^{\beta}_L\sigma_{\mu\nu}q^{\beta}_R)\\\nonumber
Q_8&=&\frac{g^2}{16\pi^2}(\bar{s}^{\alpha}_L\sigma^{\mu\nu}b^{\alpha}_R)
(\sum_q\bar{q}^{\beta}_R\sigma_{\mu\nu}q^{\beta}_L)\\\nonumber
Q_9&=&\frac{g^2}{16\pi^2}(\bar{s}^{\alpha}_L\sigma^{\mu\nu}b^{\beta}_R)
(\sum_q\bar{q}^{\beta}_L\sigma_{\mu\nu}q^{\alpha}_R)\\
Q_{10}&=&\frac{g^2}{16\pi^2}(\bar{s}^{\alpha}_L\sigma^{\mu\nu}b^{\beta}_R)
(\sum_q\bar{q}^{\beta}_R\sigma_{\mu\nu}q^{\alpha}_L)
\label{eq:ehn}
\end{eqnarray}
Wilson coefficients in  models beyond SM have been given and are listed in the following.\\
1) Model II 2HDM\\
$C_i$ i=9, 10 is given in ref.\cite{gri} and the same as that in SM when tan$\beta$ is large. For $C_{Q_i}$, 
one has\cite{dhh,hlyz} 
 \ba
\label{cq}
C_{Q_1}(m_W)&=&f_{ac} \{ \frac{sin^2(2\alpha)}{2} (1-\frac{(m_{h^0}^2-m_{H^0}^2)^2}{2 m_{h^0}^2 m_{H^0}^2})\nnb \\
&& y_t f_1(y_t) + x_t f_2(x_t,y_t)\},\nnb \\
C_{Q_2}(m_W)&=&-f_{ac} x_t f_2(x_t,y_t),\nnb \\
C_{Q_3}(m_W)&=&\frac{m_be^2}{m_{\tau}g^2}(C_{Q_1}(m_W)+C_{Q_2}(m_W)),\nnb \\
C_{Q_4}(m_W)&=&\frac{m_be^2}{m_{\tau}g^2}(C_{Q_1}(m_W)-C_{Q_2}(m_W)),\nnb \\
C_{Q_i}(m_W)&=&0, ~~~~i=5,\cdots, 10  \nnb \\
\ea
where
$$
f_{ac}=\frac{m_b m_l tan^2\beta}{4 sin^2\theta_W m_W^2},
x_t=\frac{m^2_t}{m^2_W},~~y_t=\frac{m^2_t}{m^2_{H^{\pm}}},
$$
$$
z=\frac{x}{y},  f_1(y)=\frac{1-y+ylny}{(y-1)^2},
$$
$$
f_2(x,y)=\frac{xlny}{(z-x)(x-1)}+\frac{lnz}{(z-1)(x-1)}.
$$
2) SUSY\\
There are five different sets of contributions: \\
{\it a)} the SM
contribution with exchange of $W^-$ and up-quarks; {\it b)} the charged
Higgs contribution with $H^-$ and up-quarks; {\it c)} the chargino
contribution with $\widetilde\chi^-$ and up-squarks $\tilde{u}$;
{\it d)} the gluino contribution with $\tilde g$ and down-squarks
$\tilde{d}$; and finally {\it e)} the neutralino contribution
with $\widetilde\chi^0$ and down-squarks.\\
As pointed out in ref.\cite{ber}, {\it d), e)} can be negligible in most
of region of the parameter space. 
$C_i$ i=9,10 can be found in refs.\cite{ber,got,hlyz} and $C_{Q_i}$ in refs.\cite{hy,hly,hlyz}. 
It has been shown that 
$C_{Q_i}$ i=1,2 is proportional to $\frac{m_b m_l tan^3\beta}{m_w^2}$
in some regions of the parameter space, which make the contributions from NHBs significant for l=$\mu,\tau$.

It is well known that $b \rightarrow s \gamma$ puts a very stringent constraint on the
parameter space of various models. SUSY contributions can interfere either
constructively or destructively\cite{lnwz},which is determined by the 
sign of $\mu$. Therefore, one has to consider the constraint from $B\rightarrow X_s \gamma$ in numerical
calculations of observables for $B\rightarrow X_s l^+l^-$.
The branching ratio (Br) of $B\rightarrow X_s \gamma$\cite{ahm}
\be 
2.0 \times 10^{-4} < {\cal B}(B \to X_s \gamma) < 4.5 \times 10^{-4} ~,
\label{bsgamcleo}
\ee
translates into the constraint on $C_7^{eff}$ as
\begin{equation} 
0.249 \leq \vert C_7^{eff,LLA}(\mu=4.8~\mbox{GeV}) \vert \leq 0.374~
\label{eq:c7lla}
\end{equation}
in LLA and $
0.28 \leq \vert \cse(m_B)\vert \leq 0.41~
$ in NLO. 
The constraint leads to the correlation between $C_7^{eff}$ and $C_{Q_i}$ in SUSY.

The numerical results can be summarized as follows.\\
A. In model II 2HDM for small tan$\beta$ the deviation of Br from SM is very small
and the deviation for l=$\tau$ is significant when tan$\beta$ is larger than 25 \cite{dhh,hz}. \\
B. In MSSM and mSUGRA \\
For $b\rightarrow s e^+ e^-$, there is no significant deviation from SM when tan$\beta$ is small (say, 2) and IMS is enhanced
by a factor of 50 $\%$ compared to SM in some region of the parameter space when tan$\beta$ is large (say, 30)\cite{got}.
For $b\rightarrow s \mu^+\mu^-$ and the large tan$\beta$ case, IMS can be enhanced by a factor of 100 $\%$
and FBA is significantly different from SM in some region of the parameter space due to the contributions of NHBs\cite{hly}.\\
C. In string theory and M-theory inspired models\\
For $b\rightarrow s \tau^+\tau^-$, IMS can be enhanced by a factor of 200$\%$ even 400$\%$ compared to SM in some large 
tan$\beta$ regions of the parameter space in some models because of NHB contributions\cite{hy,hlylz}.

\section{EXCLUSIVE DECAYS $B\rightarrow (K,K^*)l^+ l^-$}
For exclusive semileptonic decays of B, to make theoretical predictions, additional
knowledge of decay form factors is needed, which is related with the
calculation of hadronic transition matrix elements. Hadronic transition matrix
elements depend on the non-perturbative properties of QCD, and can only
be reliably calculated by using a nonperturbative method. The form factors for B decay into $K^{(*)}$ have been computed with
different methods such as quark models~\cite{jw}, SVZ  QCD sum rules~\cite{cfss}, light cone sum rules
(LCSRs)~ \cite{bbk}. 

IMS, FBA and LP have been given in SM\cite{see}. 
In the SM we obtain the following non-resonant branching ratios, denoted by 
${\cal{B}}_{nr}$, ($\ell=e,\mu$)\cite{alie}:
\begin{eqnarray}
&&{\cal{B}}_{nr}(B \to K \ell^{+} \ell^{-})=5.7  \cdot 10^{-7}  , \nnb \\
&&\Delta{\cal{B}}_{nr}= (^{+27}_{-15},\pm 6,^{+7}_{-6},\pm 1, \pm 
2)\%  , \nnb \\ 
&&{\cal{B}}_{nr}(B \to K \tau^{+} \tau^{-})=1.3 \cdot 10^{-7}  , \nnb \\
&& \Delta{\cal{B}}_{nr}=(^{+22}_{-6},\pm 
7,^{+4}_{-3},^{+0.4}_{-0.2}, \pm 1) \%  , \nnb \\
&& {\cal{B}}_{nr}(B \to K^* e^{+} e^{-})=2.3 \cdot 10^{-6}  ,\nnb\\ 
&&\Delta{\cal{B}}_{nr}= (^{+29}_{-17}, ^{+2}_{-9},+ 12,^{+4}_{-1},\pm 
3)\%  , \nnb\\
&& {\cal{B}}_{nr}(B \to K^* \mu^{+} \mu^{-})=1.9 \cdot 10^{-6}  , \nnb\\
&&\Delta{\cal{B}}_{nr}=(^{+26}_{-17}, \pm 6, ^{+6}_{-4}, ^{-0.7}_{+0.4}, 
\pm 2)\%  , \nnb \\ 
&&{\cal{B}}_{nr}(B \to K^* \tau^{+} \tau^{-})=1.9 \cdot 10^{-7} , \nnb\\
&&\Delta{\cal{B}}_{nr}=(^{+4}_{-8},\pm 
4,^{+13}_{-11},^{+0.6}_{-0.3},\pm 3) \%  .
 \end{eqnarray}
The first error in the $\Delta{\cal{B}}_{nr}$ consists of hadronic 
uncertainties from the form factors.
The other four errors given in the parentheses are due to 
the variations of $m_t,~\mu,~m_{b,pole}$ and $\alpha_s(m_Z)$, 
in order of appearance. In addition, there is an error of $\pm 2.5 \%$ from 
the lifetimes $\tau_B$. The largest parametric errors are
from the uncertainties of the scale $\mu$ and the top quark mass, $m_t$. 

IMS, FBA and LP have also been computed in extensions of SM\cite{alie,hlyzbk,spain} . Here we only pay attention to FBA because 
some qualitative features to discriminate new physics from SM can be seen from the observable.  For $B\rightarrow 
K^*\ell^+\ell^-$ decays FBA reads as follows\cite{alie}
\begin{eqnarray}
&&  \frac{\d \a_{\rm FB}}{\d \sh} = \nnb \\
&& \frac{G_F^2 \, \alpha^2 \, m_B^5}{2^{8} \pi^5} 
      \left| V_{ts}^\ast  V_{tb} \right|^2 \, \sh \uh(\sh)^2 \nonumber \\
&& C_{10} [  {\rm Re}(\cne) V A_1+ \frac{\mbh}{\sh} \cse (V T_2 (1-\mvh) \nnb \\
&& + A_1 T_1 (1+\mvh)) ] \;,  
\end{eqnarray}
where $A_i, T_i$ and V are relevant form factors.
The position of the zero $\sh_0$ of FBA is given by
\begin{eqnarray}
&&{\rm Re}(\cne(\sh_0)) =- \frac{\mbh}{\sh_0} \cse 
[\frac{T_2(\sh_0)}{A_1(\sh_0)} (1-\mvh) \nnb \\
&& + \frac{T_1(\sh_0)}{V(\sh_0)} (1+\mvh)],  \label{eq:fbzero}
\end{eqnarray}
which depends on the value of $m_b$, the ratio 
of the effective coefficients $\cse/{\rm Re}(\cne(\sh_0))$, and  
the ratio of the form factors shown above.
In the Large Energy Effective Theory\cite{leet}  one has a particularly simple form for the equation
determining $\sh_0$, namely\cite{alie}
\begin{equation}
{\rm Re}(\cne(\sh_0)) =- 2 \frac{\mbh}{\sh_0} \cse
\frac{1-\sh_0}{1+m_{K^*}^2 -\sh_0} \; ,
\label{eq:fbzeroleet}
\end{equation}
\begin{equation}
\label{eq:s0number}
s_0=2.88^{+0.44}_{-0.28}  \mbox{GeV}^2 
\end{equation}
 in the SM.\\
$\frac{d {\cal{A}}_{FB}}{ds}(\bksll)$
is proportional to $\ct$ and has a characteristic zero
under the condition
\begin{equation}
{\rm sign}(\cse Re(\cne)) =-1~.
\label{eq:fbsign}
\end{equation}
The condition in Eq.~(\ref{eq:fbsign}) provides a discrimination between
the SM and models having new physics. 

FBA for $\bkll$ in SUSY is given by \cite{hlyzbk}
\begin{equation}
\frac{\d \a_{\rm FB}^{K}}{\d \sh}
 = - 2 \mlh \uh(\sh){\rm Re}(\s_1 \ap^*)D^{-K}, 
\label{eq:fba}
\end{equation}
where
\be
&&D^{K}=(|\ap|^2 +|\cp|^2)
( \lam - \frac{\uh(\sh)^2}{3} ) \nnb \\
&&+ |\s_1|^2 (\sh-4 \mlh^2) 
 + |\cp|^2 4 \ml^2 (2+2 \mph^2-\sh)\nonumber\\
& & + Re( \cp \ddp^{*}) 8 \ml^2 (1-\mph^2)
+|\ddp|^2 4 \ml^2 \sh, \nonumber \\
\ee
\bea
&&  \ap(\sh)  =  \cne(\sh) \, f_+(\sh)
         + \frac{2 \mbh}{1 + \mph} \cse f_T(\sh), \; \nonumber\\
&&\s_1(\sh)  = \frac{1-\mph^2}{(\mbh-\msh)} \cqb f_0(\sh), 
\label{eq:s1}
\eea
\begin{equation}
  \uh(\sh)  = \sqrt{\lam (1-4 \frac{\mlh^2}{\sh})}  \;
\end{equation}
\begin{equation}
 \lam = 1+\hat{m}_{K,K*}^4+\sh^2-2 \sh-2 \hat{m}_{K,K*}^2(1+\sh) \;
\end{equation}
Note that the variable $\uh$ corresponds to $\theta$ through the relation $\uh = -\uh(\sh) \cos \theta$. 
It is evident from Eq. (\ref{eq:fba}, \ref{eq:s1}) that 
FBA for $\bkll$ vanishes if there are no contributions of NHBs and the contributions of NHBs can be large enough to be
observed only in SUSY and/or 2HDM with large tan$\beta$ for l=$\mu,\tau$, a non-zero FBA for $\bkll$ (l=$\mu,\tau$ ) would signal
the existence of new physics.

\section{CP VIOLATION}

In SUSY phases of soft breaking parameters provide new sources of CP violation.
The cancellation mechanism recently found makes the phases can be relatively large and experimental electric dipole 
moment (EDM) bounds for electron and neutron satisfied\cite{edm,hl1,hl2}.

The direct CP asymmetries in decay rate and backward-forward asymmetry for
 $B\rightarrow X_sl^+l^-$ and 
$\bar{B} \rightarrow X_{\bar s} l^+ l^-$ are defined as
\bea
&& A_{CP}\hspace{-12pt}^1 \hspace{12pt}(\hat{s}) = \frac{{\rm d} \Gamma
/{\rm d} \hat{s} - {\rm d}\overline{\Gamma} /{\rm d} \hat{s}}{{\rm d}
\Gamma /{\rm d} \hat{s} +{\rm d} \overline{\Gamma}/{\rm d} \hat{s}}
=\frac{D(\hat{s})-\overline{D}(\hat{s})}{D(\hat{s})+\overline{D}(\hat{s})} ,
 \nnb \\ 
&& A_{CP}\hspace{-12pt}^2 \hspace{12pt} (\hat{s}) = \frac{A(\hat{s})-
\overline{A}(\hat{s})}{A(\hat{s})+\overline{A}(\hat{s})} \nnb
\eea
\be
&& A(\hat{s}) = 3 \sqrt{ 1-\frac{4 t^2}{\hat{s}}}\frac{E(\hat{s})}{D(\hat{s})}
, \nnb 
\ee
In SUSY and the model II 2HDM\cite{hl1,hz}
\bea
&& D(\hat{s}) =  4 \big|C_7^{eff} \big|^2 (1+\frac{2}{\hat{s}}) (1+\frac
{2 t^2}{\hat{s}})\nnb\\
&&+\big|{C_9}\hspace{-4pt}^{eff} \big|^2 (2 \hat{s}+1)
(1+\frac{2 t^2}{\hat{s}}) \nnb \\
&& + \big|C_{10} \big|^2 \big[ 1+2 \hat{s}
+ (1-4 \hat{s}) \frac{2 t^2}{\hat{s}} \big] \nnb \\
&& +12 Re({C_9}\hspace{-4pt}^{eff} C_7^{eff * } (1+\frac{2 t^2}
{\hat{s}}) \nnb \\ 
&& + \frac{3}{2} \big|C_{Q_1} \big|^2 (1-\frac{4 t^2}{\hat{s}})
\hat{s} \nnb \\
&& + \frac{3}{2} \big|C_{Q_2} \big|^2 \hat{s}+ 6 Re(C_{10} C_{Q_2}^*) t
\;,
\eea
\bea 
&& E(\hat{s}) = Re( C_9\hspace{-5pt}^{eff} C_{10}\hspace{-2pt}^* \hat{s}
+2 C_7^{eff} C_{10} \hspace{-2pt}^* \nnb \\
&&+C_9\hspace{-5pt}^{eff} C_{Q_1}\hspace{-5pt}^* t
 + 2 C_7^{eff} C_{Q_1}\hspace{-5pt}^* t) \;,
\eea
\bea
&& P_N = \frac{3 \pi}{4} \sqrt{1-\frac{4 t^2}{\hat{s}}} \hat{s}^\frac{1}{2}
Im\Big[2 C_9 \hspace{-2pt}^{eff*} C_{10} t \nnb \\
&& + 4 C_{10} C_7^{eff *}
\frac{t}{\hat{s}}  + C_9\hspace{-2pt}^{eff*} C_{Q_1} \nnb \\
&& + 2 C_7^{eff *}
 C_{Q_1}+ C_{10}\hspace{-2pt}^* C_{Q_2} \Big] \Big/ D(\hat{s})
\eea

The formulas in SM can be obtained from the above formulas by taking $C_{Q_i}$=0 and $C_i$ the value in SM. The numerical
results are: 
$A_{CP}^1$ is about 0.1$\%$ in SM\cite{ks,ali2} and mSUGRA\cite{hl1} and can reach 1$\%$ in SUGRA with
non-universal gaugino masses and large tan$\beta$\cite{hl2}.  $A_{CP}^2$ is about 0.1$\%$ in mSUGRA and 
1$\%$ in SUGRA with non-universal gaugino masses respectively in large tan$\beta$  case for l=e, $\mu$. It can reach even
50$\%$ for l=$\tau$ in SUGRA with non-universal gaugino masses when tan$\beta$  is large\cite{hl2}. $P_N$ is negligible
for l=e in SM and SUSY due to  smallness of electron mass. $P_N$ is still negligible for l=$\mu$ and about 1$\%$ for
l=$\tau$ in SM\cite{ks}. It can reach 0.5$\%$ for l=$\mu$ and 5$\%$ for l=$\tau$ respectively in mSUGRA\cite{hl1}
and  6$\%$ for l=$\mu$ and 15$\%$ for l=$\tau$ respectively in SUGRA with non-universal gaugino masses when tan$\beta$
is large.

\section{CONCLUSIONS}

The following conclusions can be drawn from the above discussions.

For the inclusive decay $B\rightarrow X_s l^+l^-$, the Br in SM is smaller than experimental bound, which implies there is a room
for new physics. The deviation of the Br from SM is not significant for small tan$\beta$ and can be significant in some
large tan$\beta$ region of parameter space in SUSY. For l=$\mu, \tau$, much significant deviations can be reached 
in some large tan$\beta$ regions of parameter space in SUSY due to the NHB contributions.  And  the FBA and LP
 for $l=\mu, \tau$  are also significantly different from SM in the regions in SUSY. For the exclusive decays, Br($B\rightarrow 
K^{(*)} l^+ l^-$) in SM 
is also smaller than the data. The zero of FBA  for $\bksll$ provides a discrimination between SM and new models beyond SM.
FBA of $\bkll$ for $l=\mu, \tau$ does not vanish in some large tan$\beta$ regions of parameter space in SUSY and can be
observed for $l=\tau$ in B factories if nature chooses large tan$\beta$ and low sparticle mass spectrum. Finally, CP
violation in \bxsll  in SUGRA with non universal gaugino masses and large
tan$\beta$ can be as
 large as be observed in B factories in the future.

\end{document}